\documentclass[usenatbib]{mn2e}
\usepackage{graphicx, subfig, tabularx, bm, amssymb, natbib, amsmath, aas_macros}
\newcommand{\vmax}{V_{\rm{max}}}
\newcommand{\vpeak}{V_{\rm{peak}}}
\newcommand{\rmax}{R_{\rm{max}}}

\newcommand{\rvir}{R_{\rm{vir}}}

\newcommand{\msun}{M_{\odot}}

\newcommand{\lsun}{L_{\odot}}

\newcommand{\kpc}{{\rm kpc}}
\newcommand{\kms}{{\rm km \, s}^{-1}}

\newcommand{\lcdm}{$\Lambda$CDM}
\newcommand{\Tspace}{\rule{0pt}{2.6ex}}
\newcommand{\Bspace}{\rule[-1.2ex]{0pt}{0pt}}

\title[Satellite galaxies around the Milky Way and M31]
  {On the stark difference in satellite distributions around the Milky Way and
  Andromeda}
\author[Yniguez et al.]
{Basilio Yniguez\thanks{$\!$email: byniguez@uci.edu},  Shea Garrison-Kimmel, Michael Boylan-Kolchin\thanks{$\!$Center for Galaxy Evolution
  Fellow}, James S. Bullock\\
\noindent $\!\!$Center for Cosmology, Department of Physics and Astronomy, 
  4129 Reines Hall, University of California, Irvine, CA 92697, USA}
\begin{document}

 \pagerange{\pageref{firstpage}--\pageref{lastpage}} 
 \pubyear{2012}

\maketitle

\label{firstpage}
\begin{abstract}
  We compare spherically-averaged radial number counts of bright ($\gtrsim
  10^{5} L_{\odot}$) dwarf satellite galaxies within 400 kpc of the Milky Way (MW) and
  M31 and find that the MW satellites are much more centrally concentrated.
  Remarkably, the two satellite systems are almost identical within the central
  100 kpc, while M31 satellites outnumber MW satellites by about a factor of
  four at deprojected distances spanning 100-400 kpc. We compare the observed
  distributions to those predicted for \lcdm\ subhalos using a suite of 44
  high-resolution $\sim 10^{12} \msun$ halo zoom simulations, 22 of which are in
  pairs like the MW and M31.  We find that the radial distribution of satellites
  around M31 is fairly typical of those predicted for subhalos, while the Milky
  Way's distribution is more centrally concentrated than {\em any} of our
  simulated \lcdm\ halos.  One possible explanation is that our census of {\em
    bright} ($\gtrsim 10^{5} L_{\odot}$) MW dwarf galaxies is significantly
  incomplete beyond $\sim 100$ kpc of the Sun.  If there were $\sim 8-20$ more
  bright dwarfs orbiting undetected at 100-400 kpc distance, then the Milky
  Way's radial distribution would fall within the range expected from subhalo
  distributions and also look very much like the known M31 system.  We use our
  simulations to demonstrate that there is enough area left unexplored by the
  Sloan Digital Sky Survey and its extensions that the discovery of $\sim
10$ new bright dwarfs is not implausible given the expected range of angular
  anisotropy of subhalos in the sky.
\end{abstract}


\section{Introduction} 
\label{sec:intro}
The satellite galaxy populations around the Milky Way and M31 provide important
laboratories for galaxy formation on small scales and benchmarks for
cosmological predictions associated with dark matter halo substructure
\citep{Kravtsov_10, Bullock_2010}.  Perhaps the most famous example is the
Missing Satellites Problem, which points to a mismatch in the observed counts of
dwarf satellites compared to the predicted number of \lcdm\ dark matter subhalos
that are arguably massive enough to have formed stars \citep{Klypin_99,
  Moore_99}.  Over the last decade, discoveries of new dwarf satellite galaxies
around the Milky Way in Sloan Digital Sky Survey (SDSS; \citealt{York_00}) data
have alleviated this problem to some extent \citep{Willman_05, Zucker_06a,
  Zucker_06b, Belokurov_06, Grillmair_06, Walsh_07, Irwin_07}.  Specifically, a
new population of {\em ultra-faint} dwarfs ($L \simeq 10^{2} - 10^{4}\,\lsun$)
have been discovered that that pile up along the radial detection limit of the
SDSS \citep{Koposov_08, Walsh_09}.  Completeness corrections suggest that there
could be as many as a few hundred {\em ultra-faint} dwarfs within the virial
radius of the Milky Way \citep{Tollerud_08, Walsh_09}.

Given that our Galactic satellite census suffers from radially-biased
incompleteness effects, M31 and its satellite system provides a particularly
useful comparison set.  The observational biases with M31 satellite searches are
qualitatively different from those faced in Milky Way.  In particular, we view
M31 from the outside, yet are close enough to detect fairly low luminosity dwarf
spheroidal galaxies there.  The Pan-Andromeda Archeological Survey (PAndAS;
\citealt{McConnachie_09}) provides a uniform search for dwarfs around M31, and has
thus far discovered about half of the known M31 satellites.  Their coverage is
complete to within a projected distance of 150 kpc and outer M31 halo coverage
is ongoing \citep{Richardson_11}.  Interestingly, the current M31 dwarf count,
which is almost certainly incomplete beyond 150 kpc, is larger than the Milky
Way count by about a factor of two at fixed luminosity (see below).  This
difference is striking given that the two spirals have the same luminosity to
within $\sim 30 \%$ \citep{vandenbergh_00}.
 
Beyond the question of overall counts, another useful comparison concerns the
relative radial distribution of satellite galaxies.  The Milky Way satellites
are known to be more centrally concentrated around the Galaxy than would be
expected for subhalos in dissipationless \lcdm\ simulations
\citep{Moore_01,Willman_04,Maccio_2010}.  This basic mismatch has been used as a
potential testing ground for ideas to solve the Missing Satellites Problem.  For
example, early-forming halos that formed before reionization tend to be more
centrally concentrated than the overall population \citep{Bullock_00,Moore_01}.
The problem with this explanation is that the mismatch between naive theory and
observation appears to be much less severe for M31.
 
\citet{Willman_04} were the first to emphasize that known M31 satellite galaxy
population was somewhat more extended than that of the Milky Way, and suggested
that the disagreement could be explained by incompleteness in the outer Milky
Way halo, possibly as a result of the difficulty in detecting low-surface
brightness systems at low Galactic latitude.  Importantly, at the time of their
work, all of the known satellites of M31 and the Milky Way were ``bright" by
today's standards, with luminosities $L > 10^{5}\,\lsun$.  They emphasized
that this possible incompleteness in (bright) satellites provided motivation for
searches for new dwarfs via resolved-stars in the Sloan Digital Sky Survey.  As
already mentioned, these searches have proven immensely successful at finding
new {\em faint} dwarfs, but new bright dwarfs of the type known before 2004 have
remained sparse: all but one of the $\sim 15$ new dwarfs that have been
discovered since 2004 are less luminous than $10^{5}\,\lsun$ and none are
more luminous than $3 \times 10^5 \lsun$.  Over the same period, 20 new M31
satellite galaxies have been discovered, 14 of which are {\em more} luminous
than $10^{5}\,\lsun$.

In this paper we reexamine the radial distributions of M31 satellites and Milky
Way satellites.  We restrict ourselves to the brightest systems: bright enough
that we could possibly assume our census of them complete within the Milky Way virial radius. 
 Given the new discoveries of mainly
bright satellites around M31, we expect to see the difference in their radial
populations to be even more discrepant than previously quantified
\citep{McConnachie_06}.  This discrepancy may grow yet with the discovery of
additional satellites around M31 as PAndAS coverage expands
\citep{Richardson_11}.
  
The outline of this paper is as follows: in Section 2 we describe the data sets
of the MW, M31 and our suite of \lcdm\ simulations.  Section 3 contains the results
of our comparison, Section 4 discusses the possible effects of incompleteness
in the satellite populations and in Section 5 we conclude.

\begin{center}
 \begin{table*}
  \begin{tabular}{ | l | c | c | c || l | c | c | c | }
\hline
\textbf{Name}& $\bm{d_{\rm M31}}$ [kpc]&$\bm{L_V \,[\lsun]}$ &\textbf{Discovered} & 
\textbf{Name}& $\bm{d_{\rm MW}}$ [kpc]&$\bm{L_V \,[\lsun]}$ &\textbf{Discovered} \\
\hline
M31 & - & $2.6 \times 10^{10}$ & - & 			 Milky Way & - & $2.0 \times 10^{10}$ & -  \\
\hline
M32 \Tspace \Bspace & $      23^{      +45}_{     -17}		$ & $ 2.9\times10^{8}$ &   1749&                   LMC  & $49\pm   3$ &$ 2.2\times 10^{9} $&-\\

M33 \Tspace \Bspace & $     209^{       +7}_{      -5}		$ & $ 2.8\times10^9$ &   1764&                     SMC  & $58\pm   4$ &$ 5.9\times 10^{8} $&-\\
                                                                                                                   
NGC205 \Tspace \Bspace & $      42^{      +27}_{     -25}	$ & $ 3.7\times10^{8}$ &   1783&                  Sculptor & $79 \pm4$ &$7.1\times 10^{5} $&1937\\
                                                                                                                   
NGC185 \Tspace \Bspace & $     185^{      +16}_{     -16}	$ & $ 1.8\times10^{8}$ &   1787&                   Fornax  & $ 140   \pm8$ &$ 1.5\times 10^{7} $&1938\\
                                                                                                                   
NGC147 \Tspace \Bspace & $     120^{      +11}_{     -10}	$ & $ 1.4\times10^{8}$ &   1829&                    LeoI  & $ 254   \pm   30$ &$ 4.9\times 10^{6 }$&1950\\
                                                                                                                   
IC10 \Tspace \Bspace & $     252^{      +11}_{      -5}		$ & $ 8.6\times10^{7}$ &   1888&                  LeoII  & $ 208   \pm   12$ &$9.4\times 10^{5} $&1950\\
                                                                                                                   
AndI \Tspace \Bspace & $      71^{+      14}_{     -13}		$ & $ 4.5\times10^{6}$ &   1971&        Ursa Minor  & $68\pm   3$ &$ 2.8 \times 10^{5} $&1954\\
                                                                                                                   
AndII \Tspace \Bspace & $     198^{+      11}_{     -10}		$ & $ 9.4\times10^{6}$ &   1974&           Draco  & $76\pm   5$ & $ 2.8\times 10^{5} $&1954\\
                                                                                                                   
AndIII \Tspace \Bspace & $      88^{+      17}_{     -10}	$ & $ 1.0\times10^{6}$ &   1974&                   Carina & $ 103   \pm5$ &   $   4.9\times 10^{5} $& 1977\\
                                                                                                                   
Pisces \Tspace \Bspace & $     268^{       +4}_{      -2}		$ & $      9.4\times10^{5}$ &   1976&      Sextans  & $89 \pm  4$ &$5.4\times 10^{5} $&1990\\
                                                                                                                   
AndV \Tspace \Bspace & $     115^{+       8}_{      -4}		$ & $      7.1\times10^{5}$ &   1998&               Sagittarius  & $20 \pm  4$ & $8.6\times 10^{7} $&1994\\
                                                                                                                   
AndVI \Tspace \Bspace & $     269^{+       5}_{      -3}		$ & $ 3.4\times10^{6}$ &   1999&           & & & \\

AndVII \Tspace \Bspace & $     219^{       +5}_{      - 0}	$ & $ 1.8\times10^{7}$ &   1999&                   & & & \\
\hline
\hline
AndIX \Tspace \Bspace & $     186^{      +23}_{     -87}		$ & $      1.\times10^{5}$ &   2004&        Ursa Major I  & $ 105   \pm   10$ &$1.4\times 10^{4} $&2005\\   
                                                                                                                   
AndX \Tspace \Bspace & $     126^{      +33}_{     -18}		$ & $      1.5\times10^{5}$ &   2006&          Bootes I  & $64 \pm  6$ &$2.8\times 10^{4} $&2006\\
                                                                                                                   
AndXI \Tspace \Bspace & $     103^{      +54}_{      -1}		$ & $      4.9\times10^{4}$ &   2006&     Hercules  & $ 145 \pm 13$ &$3.7\times 10^{4} $&2006\\

AndXII \Tspace \Bspace & $     178^{      +35}_{     -73}	$ & $      3.1\times10^{4}$ &   2006&             Canes Ven. I  & $ 220 \pm 20$ &$2.4\times 10^{5} $&2006\\

AndXIII \Tspace \Bspace & $     116^{      +88}_{      -1}	$ & $      4.1\times10^{4}$ &   2006&               &  &  &  \\
                                                                                                                  
AndXIV \Tspace \Bspace & $     161^{      +59}_{      -2}	$ & $      1.8\times10^{5}$ &   2007&             &  &  &  \\
                                                                                                                  
AndXV \Tspace \Bspace & $     178^{      +31}_{     -60}		$ & $      4.9\times10^{5}$ &   2007&     &  &  &  \\
                                                                                                                  
AndXVI \Tspace \Bspace & $     323^{      +27}_{     -40}	$ & $      4.1\times10^{5}$ &   2007&             &  &  &  \\
                                                                                                                  
AndXVII \Tspace \Bspace & $      70^{      +24}_{     -23}	$ & $      2.1\times10^{5}$ &   2008&            &  &  &  \\
                                                                                                                  
AndXIX \Tspace \Bspace & $     114^{      +33}_{      -9}	$ & $      4.5\times10^{5}$ &   2008&            &  &  &  \\
                                                                                                                  
AndXX \Tspace \Bspace & $     130^{      +21}_{      -4}		$ & $      2.8\times10^{4}$ &   2008&    &  &  &  \\
                                                                                                                  
AndXXI \Tspace \Bspace & $     134^{      +11}_{      -8}	$ & $      7.8\times10^{5}$ &   2009&            &  &  &  \\
                                                                                                                  
AndXXII \Tspace \Bspace & $     270^{      +21}_{     -56}	$ & $      5.4\times10^{4}$ &   2009&            &  &  &  \\
                                                                                                                  
AndXXIII \Tspace \Bspace & $     129^{       +6}_{      -2}	$ & $   	 1.0\times10^{6}$ &   2011&      &  &  &  \\
                                                                                                                  
AndXXIV \Tspace \Bspace & $     166^{      +22}_{     -28}	$ & $      9.4\times10^{5}$ &   2011&            &  &  &  \\
                                                                                                                  
AndXXV \Tspace \Bspace & $      93^{      +46}_{      -8}	$ & $      6.5\times10^{5}$ &   2011&            &  &  &  \\
                                                                                                                  
AndXXVI \Tspace \Bspace & $     104^{     +116}_{      -2}	$ & $      5.9\times10^{4}$ &   2011&            &  &  &  \\
                                                                                                                  
AndXXVII \Tspace \Bspace & $     478^{      +41}_{    -420}	$ & $      1.2\times10^{5}$ &   2011&            &  &  &  \\
                                                                                                                  
AndXXVIII \Tspace \Bspace & $     369^{      +17}_{      -2}	$ & $      2.1\times10^{5}$ &   2011&            &  &  &  \\
                                                                                                                  
AndXXIX \Tspace \Bspace & $     188^{      +25}_{      -3}	$ & $      1.8\times10^{5}$ &   2011&            &  &  &  \\
\hline
  \end{tabular}
  \caption{Satellite galaxies of M31 (left) and Milky Way (right) used in this
    work, listed in order of their discovery date.  The horizontal band marks
    the transition from the ``classical" era of discovery to the
    post-SDSS/PAndAS era.  With a deprojected distance greater than 400 kpc
    (with relatively small uncertainty), And
    XVIII may not be a bound satellite of M31; it is therefore not listed here
    or included in any figures.  For completeness, we list all satellites
    brighter than $L_V = 10^4 \lsun$ here, though we only present data for
    systems fainter than  $L_V = 10^5\lsun$ in Figure~\ref{fig:lum_func}.  We
    use the more 
    conservative cut at $L_V = 10^5 \lsun$ for subsequent figures and radial
    comparisons.  Luminosities are taken from \citet{Watkins_12} and
    \citet{McConnachie_12} while the host-satellite separations and associated
    errors were computed from the heliocentric distances quoted by those
    authors.} 
\end{table*}
\end{center}

\section{Data}
\label{sec:data}
\subsection{Satellites of the MW and M31}
\label{subsec:satellites}
The M31 and Milky Way satellite data sets we examine in this paper are
summarized in Table 1, with data taken from the review by
\citet{McConnachie_12}.  Galaxies are listed in the order they were found in
order to emphasize the great explosion of discoveries in the post SDSS/PAndAS
era (horizontal break).  We include only systems that could be within 400 kpc
physical separation of each host, given published distance uncertainties.

The Milky Way satellites listed in Table 1 are restricted to those brighter than
the faintest known M31 dwarfs ($ > 10^4 \lsun$), 15 in all.  Twelve of these are
brighter than $10^5\lsun$, the approximate completeness limit of SDSS at 400 kpc
and of PAndAS for M31 satellite discovery \citep{Richardson_11}.  Only one of
the new dwarfs discovered in the $\sim 1/3$ of the sky covered by the SDSS is
brighter than this: Canes Venatici I at $220$ kpc distance.  This has provided some
motivation for assuming that the Milky Way census of {\em bright} dwarfs is
reasonably complete out to the virial extent of the Milky Way ($\sim 300$ kpc),
though it will be impossible to quantify this completeness with certainty until
uniform, full-sky, resolved-star searches have been completed.
	
The M31 data set includes a total of 32 galaxies, more than half of which were
discovered in the last six years.  The census of dwarfs in the vicinity of M31
is well-understood within a projected radius of about 150 kpc of the M31 center
\citep{Richardson_11}.  Although the faintest known M31 dwarfs known have
luminosities down to $\sim 10^{4} \lsun$, the PAndAS survey's limit for
efficient detection is closer to $L_V \sim 10^{5}\lsun$ \citep{Richardson_11}.
This completeness limit applies only to the region within 150 kpc of the M31
disk.  Of the 32 M31 satellites listed, 26 are brighter than $10^5\lsun$.  We
will use this luminosity as a characteristic luminosity for comparison in this
work.  We have not included And VIII and And XVIII in our comparison set.
\citet{Merrett_06} have questioned And VIII's classification as a galaxy because
of tidal features.  And XVIII has a very high deprojected separation from the
center of M31, falling outside of our 400 kpc region of consideration.

The deprojected distances and errors in Table 1 were derived from the
heliocentric distances and errors quoted in \citet{McConnachie_12}.  The
separation between M31 and one of its satellites is given by
\begin{equation}
r_{M31} = \sqrt{R^2 + d^2 -2 \,R \, d \cos{\theta}}\,,
\end{equation} 
where R is the heliocentric distance of M31, d is the heliocentric distance of
the satellite and $\theta$ is the angular separation between the two.  We
sampled the heliocentric distance of M31 and of each satellite 1000 times,
drawing from a Gaussian based on quoted line-of-sight errors to obtain the error
limits of $r_{M31}$, which are not always Gaussian-like themselves.

%%%%%%%%%%%%%%%%%%%%%%%%%%%%%%%%%%%%%%%%%%%%%%%%%%%%%%%%%%%%%%
\begin{figure*}
\begin{center}
 \includegraphics[width=0.9\textwidth, viewport= -1 20 515 255]{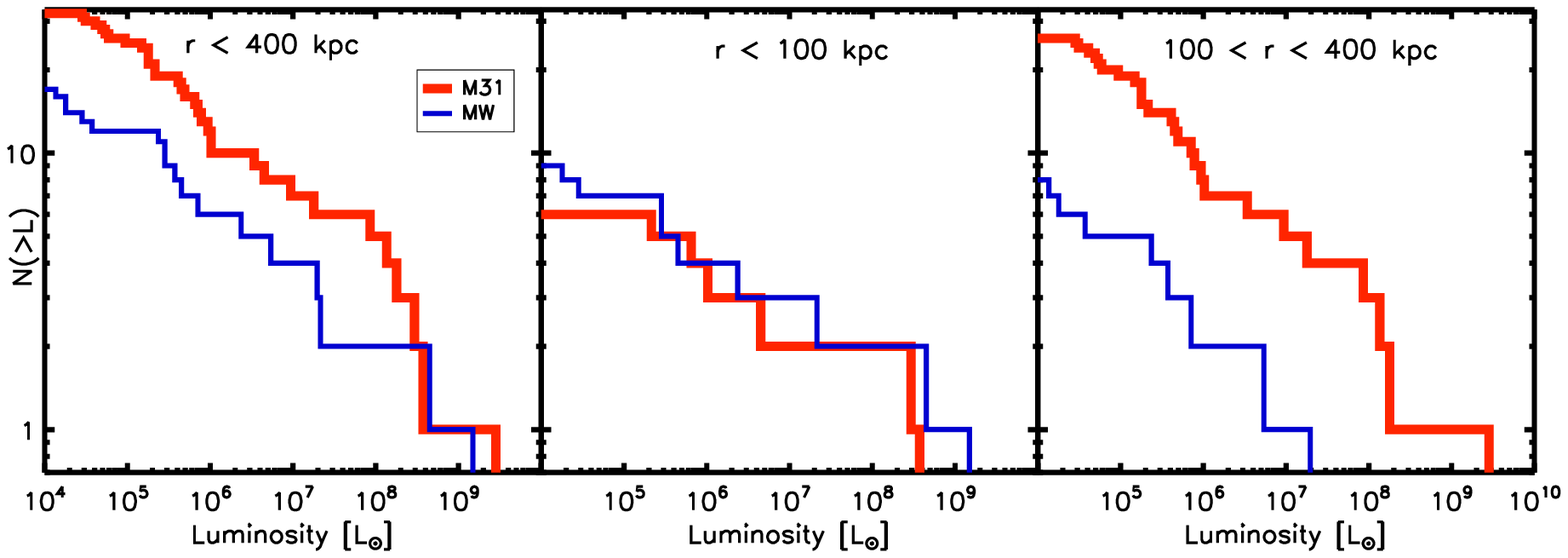}
 \vspace{-.7in}
 \caption{Satellite galaxy luminosity functions of the MW (blue) and M31 (red)
   for all galaxies within $400$ kpc of either host (left); those within 100 kpc
   of either host (middle); and those in the radial range 100-400 kpc
   (right). While the luminosity functions within 100 kpc are very similar, M31
   has approximately four times as many satellites at fixed luminosity in the
   100-400 kpc range.
 \label{fig:lum_func}
}
\end{center}
\end{figure*}
%%%%%%%%%%%%%%%%%%%%%%%%%%%%%%%%%%%%%%%%%%%%%%%%%%%%%%%%%%%%%% 
%%%%%%%%%%%%%%%%%%%%%%%%%%%%%%%%%%%%%%%%%%%%%%%%%%%%%%%%%%%%%%

\subsection{Simulations}
\label{subsec:simulations}
We use a total of 44 Milky-Way size halos simulated as part of the ELVIS
project, which is described in detail in Garrison-Kimmel et al. (in
preparation); here we summarize the relevant properties.  ELVIS is a suite of
collisionless zoom-in simulations designed to study the Local Group. It consists
of twenty-two halos in paired systems that were chosen to resemble the M31 and
Milky Way in mass and phase-space configuration in addition to twenty-two halos
that are isolated, mass-matched analogues, which serve as a control sample to
examine any trends due to the paired nature of the Local Group system.  The host
dark matter halos have virial masses between $1.0 - 2.8 \times 10^{12} \msun$ and
associated virial radii $\rvir = 265 - 370$ kpc, with maximum circular
velocities that range from $\vmax = 155 - 225\,\kms$.

All ELVIS halos were simulated in a \lcdm\ cosmology with parameters based on
WMAP7 results \citep{Larson_11}: $\sigma_8 = 0.801$, $\Omega_m = 0.266$,
$\Omega_\Lambda = 0.734$, $n_s = 0.963$, and $h = 0.71$.  The simulations were
run with identical mass resolution (particle mass $m_p = 1.9\times10^5\msun$);
likewise, all simulations were run with a Plummer-equivalent force softening of
$140\,\mathrm{pc}$, which was comoving until $z = 9$, after which a physical
softening length was imposed.  All simulations are uncontaminated by
low-resolution particles to at least 900 kpc from the center of each host.  Halo
substructure was identified with \texttt{Rockstar} \citep{Behroozi_12a}, and
merger trees were built using \texttt{consistent-trees} \citep{Behroozi_12b}.
At this resolution, we are complete to subhalos with current maximum circular
velocity $\vmax > 8\,\kms$ and to those with $\vpeak > 10\,\kms$, where $\vpeak$
is the largest circular velocity ever achieved by the main branch of a subhalo
progenitor.

Additionally, two of our host isolated halos were resimulated at an even higher
resolution with a particle mass of $m_p = 2.35 \times 10^4 \msun$ and a force
resolution of $70\,{\rm pc}$, comparable to the resolution of the level 2
Aquarius simulations \citep{springel2008} and the Via Lactea I simulation
\citep{diemand2007}.  The velocity functions and radial distributions of the two
higher resolution host halos are nearly identical to those of the lower
resolution counterparts to $\vpeak$ values well below those used for selecting
subhalos in this work.

In our fiducial radial distribution comparisons, we select the 30 subhalos with
the highest $\vpeak$ within 400 kpc of each host.  This choice is motivated by
the fact that M31 has approximately 30 satellite galaxies within this radius;
this is also the approximate distance to which SDSS is complete to
$L>10^5\,\lsun$ dwarfs.  In practice, this amounts to selecting halos with
$\vpeak$ values greater than $20\,\kms$ to $33\,\kms$, depending on the host.
The overall shapes of the distributions are fairly insensitive to this
specific choice.  For example, when we chose the 100 most massive subhalos, or
when we use a fixed $\vpeak$ cut, we find that the radial distributions of
subhalos do not change significantly. In the Appendix, we explore the spatial
distributions of subhalos with the 30 largest values of $\vmax$ and those with
the 30 largest values of $z_{\rm peak}$ (the redshift at which $\vpeak$ is
attained). For the latter case, we restrict ourselves to subhalos with $\vmax >
10\,\kms$.

\section{Results}
\label{sec:results}
\subsection{The MW and M31 satellite systems compared}
\label{subsec:mw_vs_m31}
The satellite luminosity functions of the MW and M31 are shown in
Figure~\ref{fig:lum_func}. The left panel includes all satellites within a
deprojected radius of 400 kpc of either host.\footnote{In this figure we have
  used the best-fit radial distances of satellites listed in Table 1.}
Considering this full sample, M31 has approximately a factor of two more
satellites at fixed luminosity. Intriguingly, however, the two systems have
luminosity functions that are essentially identical when only satellites within
100 kpc are considered (middle panel). The difference in satellite luminosity
functions between the two systems is entirely attributable to differences at
large radial separations (100 - 400 kpc), as shown in the right panel of Figure~\ref{fig:lum_func}.
 While M31 has approximately 20 satellites brighter than
$10^5 \lsun$ in this radial range, the Milky Way has only 5. We note, however,
that the shape of the luminosity function remains very similar at these large
radii, even though the amplitude differs by a factor of $\sim 4$.

A complementary demonstration of the difference between the satellite systems of
the two Local Group giants is given in Figure~\ref{fig:n_cum_obs}, which compares
the cumulative radial count of satellites brighter than $10^5 \lsun$ around each
host. Solid lines show median distributions and the dashed lines show 68\%
uncertainties derived using Monte Carlo realizations of the deprojected
distances and errors listed in Table 1. M31's satellite population is
significantly more extended than that of the MW, almost entirely due to
differences at \textgreater~100 kpc.  

In the context of observational completeness, it is remarkable that M31 has so
many more satellites at a large galactocentric distances. PAndAS has provided a
uniform search for (bright) dwarf galaxies only within a projected distance of
150 kpc of the Andromeda galaxy \citep{Richardson_11}, so it
would not be surprising if there are even more M31 dwarfs waiting to be
discovered beyond 150 kpc -- a region marked by the vertical dashed line in
Figure~\ref{fig:n_cum_obs}. Any such discoveries would only enhance the
discrepancy between the MW and M31.

One way the difference in cumulative radial distributions might be reconciled is
if our M31 census were incomplete at {\em small} projected radius, perhaps
because of the difficulty in detecting low surface brightness dwarfs in the
vicinity of M31's disk.  Interestingly, of the M31 dwarfs less luminous than
$10^6 \lsun$, And IX is the closest at a projected distance of R = 36 kpc.
Meanwhile, there are 2 (or 3, if And VIII is included) very luminous satellites
that are within this radius.  If there were a few more low-luminosity dwarfs
hiding in glare of the M31 disk, this would act to shift the M31 satellite
distribution towards a more concentrated shape, more in line with that of the
MW.  However, discoveries of this kind would further exacerbate the overall
difference in {\em normalization} between the two satellite luminosity
functions.  Moreover, additional small-R discoveries around M31 would also drive
the M31 radial distribution away from the expectations for subhalos derived from
our \lcdm\ simulations.  As we demonstrate in the next section, the distribution
of presently-known M31 satellites is much more in line with subhalo
distributions derived from our LCDM simulations than the known satellite
distribution around the Milky Way.

\subsection{Comparison to Simulations} 
\label{subsec:sim_comparison}
%%%%%%%%%%%%%%%%%%%%%%%%%%%%%%%%%%%%%%%%%%%%%%%%%%%%%%%%%%%%%% 
\begin{figure}
 \hspace{-15mm}
 \includegraphics[width=0.65\textwidth]{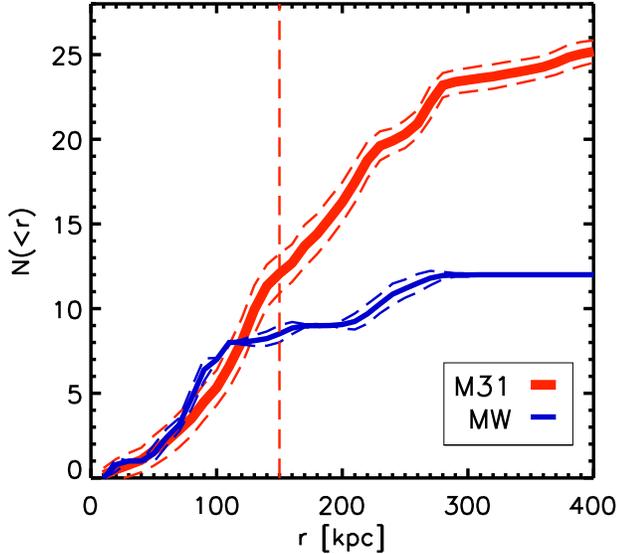}
 \caption{Cumulative number of satellites with $L > 10^5\lsun$ within a
   deprojected distance $r$ from either the MW or M31.  The solid and dashed
   lines show the median and 68\% confidence intervals distributions implied by
   current observational uncertainties. The spatial distributions are
   similar within $\sim$ 100 kpc, while the M31 satellite system is
   significantly more extended than that of the MW outside of this radius. This
   is in spite of the fact that the M31 system is almost certainly incomplete
   at distances on the right of the vertical dashed line, which marks the edge of the region
   within which the PAndAS project has provided a uniform search for M31 dwarfs.
   \label{fig:n_cum_obs}}
\end{figure}
%%%%%%%%%%%%%%%%%%%%%%%%%%%%%%%%%%%%%%%%%%%%%%%%%%%%%%%%%%%%%% 
%%%%%%%%%%%%%%%%%%%%%%%%%%%%%%%%%%%%%%%%%%%%%%%%%%%%%%%%%%%%%%
\begin{figure}
\hspace{-15mm}
 \includegraphics[width=0.65\textwidth]{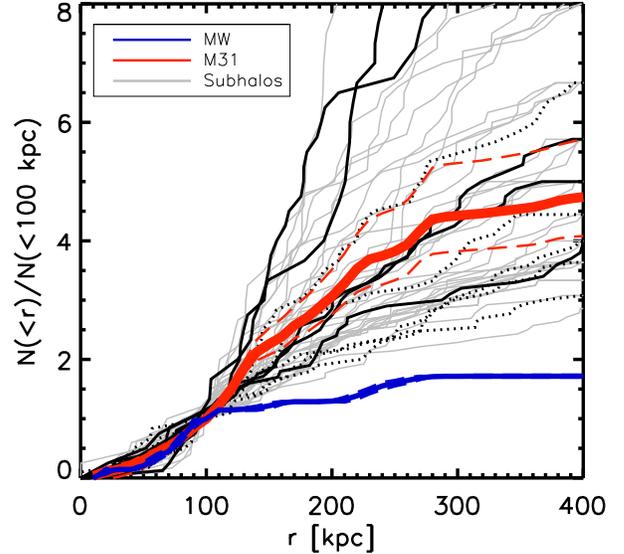}
 \caption{The cumulative number of satellites within radius $r$, normalized to
   the count within $r=100$ kpc separation.  The solid (dashed) blue and red
   lines show the median (68\% spread) profiles for the MW and M31 satellite
   systems while the grey and black lines show the profiles of subhalo
   distributions derived from 44 $\Lambda$CDM simulations.  The subhalo
   distributions were made from the 30 subhalos with the highest $\vpeak$ from
   each simulation.  The Milky Way's distribution is more centrally concentrated
   than any of the simulated systems, while M31 looks fairly typical.   For
   reference, the five solid black lines show subhalo profiles for the five
   highest virial mass halos in our sample ($2.7 - 2.8 \times 10^{12} \msun$)
   and the black dotted lines lines are from our five lowest mass halos ($1.0 -
   1.2 \times 10^{12} \msun$).  While there is some tendency for the highest
   mass systems to have more extended distributions (when measured this way),
   the correlation is weak and shows substantial scatter.
   \label{fig:n_cum_subs}
}
\end{figure}
%%%%%%%%%%%%%%%%%%%%%%%%%%%%%%%%%%%%%%%%%%%%%%%%%%%%%%%%%%%%%%

%%%%%%%%%%%%%%%%%%%%%%%%%%%%%%%%%%%%%%%%%%%%%%%%%%%%%%%%%%%%%% 
\begin{figure}
\hspace{-16mm}
 \includegraphics[width = 0.65\textwidth]{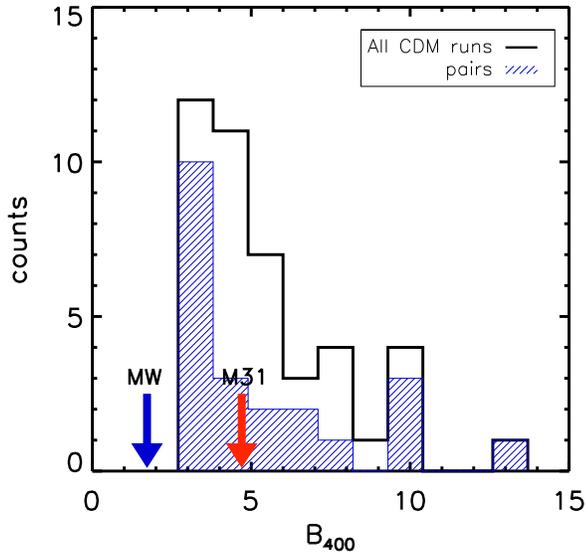}
 \caption{The distribution of subhalo radial distribution shapes --- $B_{400} \equiv
   N(<400 {\rm kpc})/N(<100 {\rm kpc})$ --- for all of our simulated halos (solid
   line) with the observed values for the MW and M31 indicated by arrows.  The
   shaded histogram shows the subset of simulated halos that are in Local Group
   like pairs.  The MW is clearly an outlier from the simulated
   distributions in that its satellite galaxy population is much more
   concentrated; M31 is more typical of the simulated systems. 
   \label{fig:b400}}
\end{figure}
%%%%%%%%%%%%%%%%%%%%%%%%%%%%%%%%%%%%%%%%%%%%%%%%%%%%%%%%%%%%%% 
Since the Milky Way and M31 have such different radial distributions of
satellites, it is natural to ask which (if either) is more in line with
theoretical expectations. We expect a correspondence between subhalos with large
values of $\vpeak$ and luminous satellites (e.g., \citealt{Bullock_2010,
  Kravtsov_10}). Figure~\ref{fig:n_cum_subs} therefore compares the observed
radial distributions of satellites of the MW (blue) and M31 (red) to those of
the 30 top $\vpeak$ subhalos from our 44 \lcdm\ simulation (gray lines); for
each data set, the radial profile is normalized to the cumulative count within
100 kpc. As in Figure~\ref{fig:n_cum_obs}, the dashed colored lines show
uncertainties associated with deprojection for the observational data.

Although there is significant scatter in the number of satellites at large radii
($\gtrsim 300 \,\kpc$) in the simulations, it is clear that the M31 system is
consistent with the profiles expected from subhalo distributions while the MW is
more centrally concentrated than any of the simulated systems. A typical
simulated halo has 4-5 times as many satellites within 300 kpc as within 100
kpc, while the corresponding number for the Milky Way is only 1.5. 

The solid (dotted) black lines in the Figure correspond to the five highest
(lowest) mass host halos in our sample. While two of the five most massive halos also
have subhalo populations that are among the most extended of our sample, as
might be expected based on their large virial radii, three of the five most
massive hosts have subhalo populations that are either fairly typical or even
somewhat more concentrated than average. Any trend with host halo mass is fairly
weak and is overwhelmed by scatter at fixed mass (this was also seen by
\citealt{Wang_13}).

It is useful to have a parameterization of the degree of concentration of a
halo's satellite population. For this we define a parameter $B_{400}$ to be
the ratio of the number of satellites within $400$ kpc to the number within
$100$ kpc:
\begin{equation}
B_{400} \equiv N(<400\, {\rm kpc})/N(<100\, {\rm kpc}).  
\end{equation}
$B_{400}$ is then a measure of the ``puffiness" of the subhalo
population. The distribution of $B_{400}$ values for all of our simulated host systems is
 shown by the solid histogram in Figure~\ref{fig:b400}, while the measured
values for the MW and M31 are indicated by arrows. As expected, we see that the
satellite distribution of the MW is more concentrated than 
\textit{any} of the simulated subhalo systems, while M31 is fairly typical of what is seen
in the simulations. The shaded blue histogram shows the subset of our simulated
halos that are paired, Local Group analogs; these halos are fairly similar to the
isolated, mass-matched analogs in their $B_{400}$ distributions.  As a result, the fact
that M31 and the MW are paired should not significantly skew their satellite
radial distributions compared to field systems (at least when spherically
averaged; we will explore angular anisotropy in more detail in a future paper).

Even though the MW is an outlier relative to all of our simulated systems, one
might posit that the high concentration of the Milky Way's satellite population
is attributable to some physical process. However, we were unable to find any
correlation between $B_{400}$ and halo concentration, mass accretion history, or virial mass. In the Appendix, we show
explicitly that the radial subhalo distributions are not tightly correlated with
the bulk infall time distributions of satellites.  Given that we are unable to
provide a natural explanation for why the MW system is highly anomalous compared
to predicted subhalo distributions (while M31 is not), in the next section we
explore the possibility that the census of bright MW satellites may be
incomplete at radii beyond 100 kpc.

In the Appendix we explore a few other options for selecting subhalos and the
associated implications for radial distribution.  For most reasonable choices,
the selection on $\vpeak$ we use for our fiducial analysis produces a more
centrally concentrated subhalo distribution than those with the highest $\vmax$.
However, we show that by sub-selecting $\vmax > 10\,\kms$ subhalos that reach
their $\vpeak$ values the earliest (i.e. were accreted into the host halo the
earliest) we can produce subhalo distributions that are more centrally
concentrated (see also \citealt{Wang_13}), and more in line with the observed
radial distribution of the Milky Way.  However, this model is not very well
motivated as a physical selector for luminous subhalos.  As shown in the
Appendix, it effectively picks out $\sim 10\%$ of the subhalos that have $\vpeak
< 30 \kms$, and ignores an equal number of systems with $\vpeak > 30 \kms$.
There is no astrophysical reason to expect these most massive progenitors to be
dark.

\subsection{Are there undiscovered bright satellites around the Milky Way?}
\label{subsec:bright_mw}
The Milky Way satellite census may still be incomplete at large radii, even to
satellites similar in luminosity to the classical dwarfs ($L > 10^{5} \lsun$).
The only truly uniform survey with well-understood completeness limits is
SDSS/SEGUE, which has covered approximately $1/3$ of the sky.  One bright dwarf
($L>10^{5}\,\lsun$), Canes Venatici I, was discovered in this area
\citep{Zucker_06b}.  Is it therefore worthwhile to ask the question: how many
new Milky Way dwarfs could plausibly be undiscovered?  We investigate this question 
first in terms of sky coverage and angular anisotropy, then in terms of radial distributions.

%%%%%%%%%%%%%%%%%%%%%%%%%%%%%%%%%%%%%%%%%%%%%%%%%%%%%%%%%%%%%%
\begin{figure*}
 %\hspace{-15mm}
 \includegraphics[width = 0.5\textwidth]{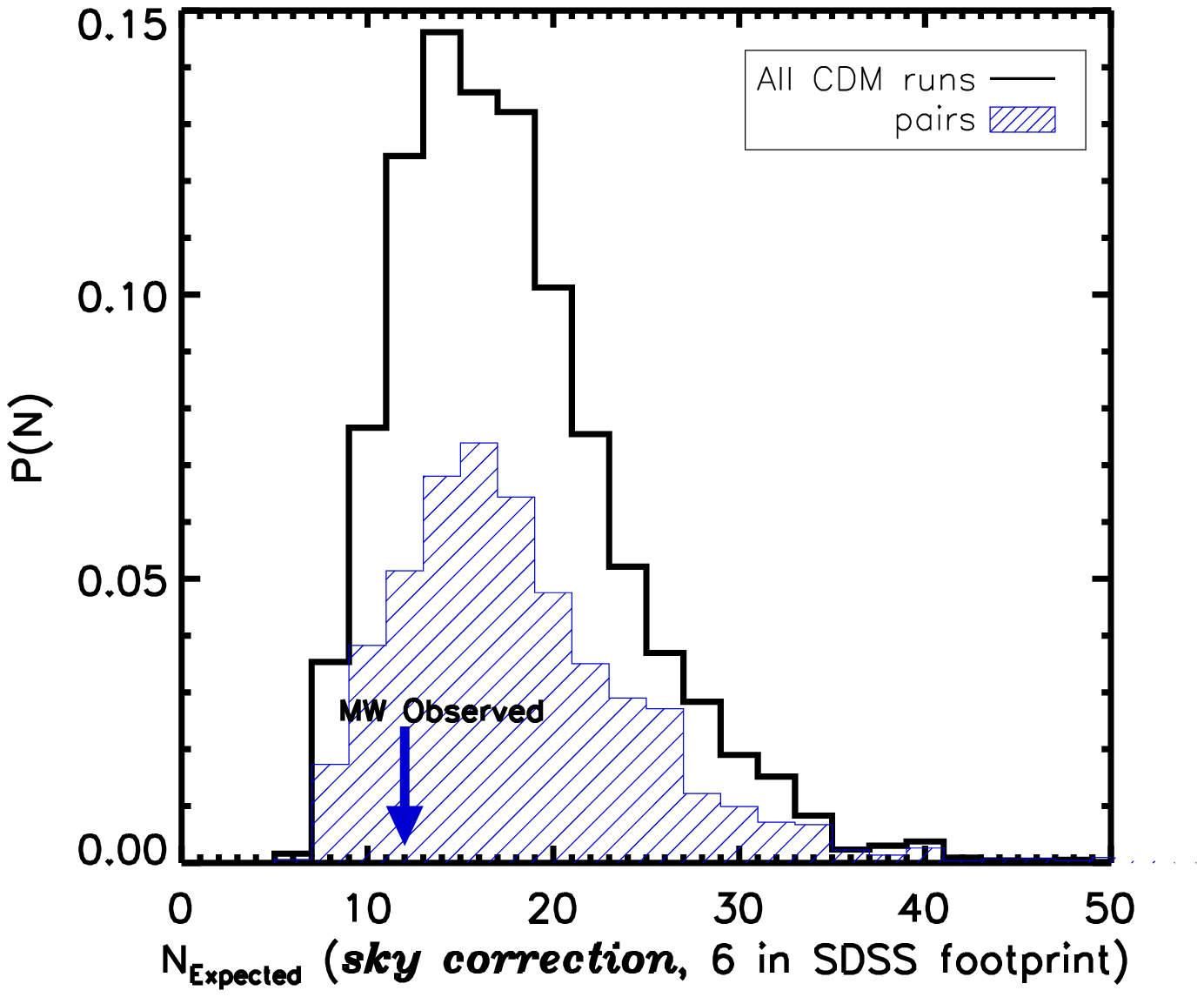}
 \hspace{-19mm}
 \includegraphics[width = 0.5\textwidth]{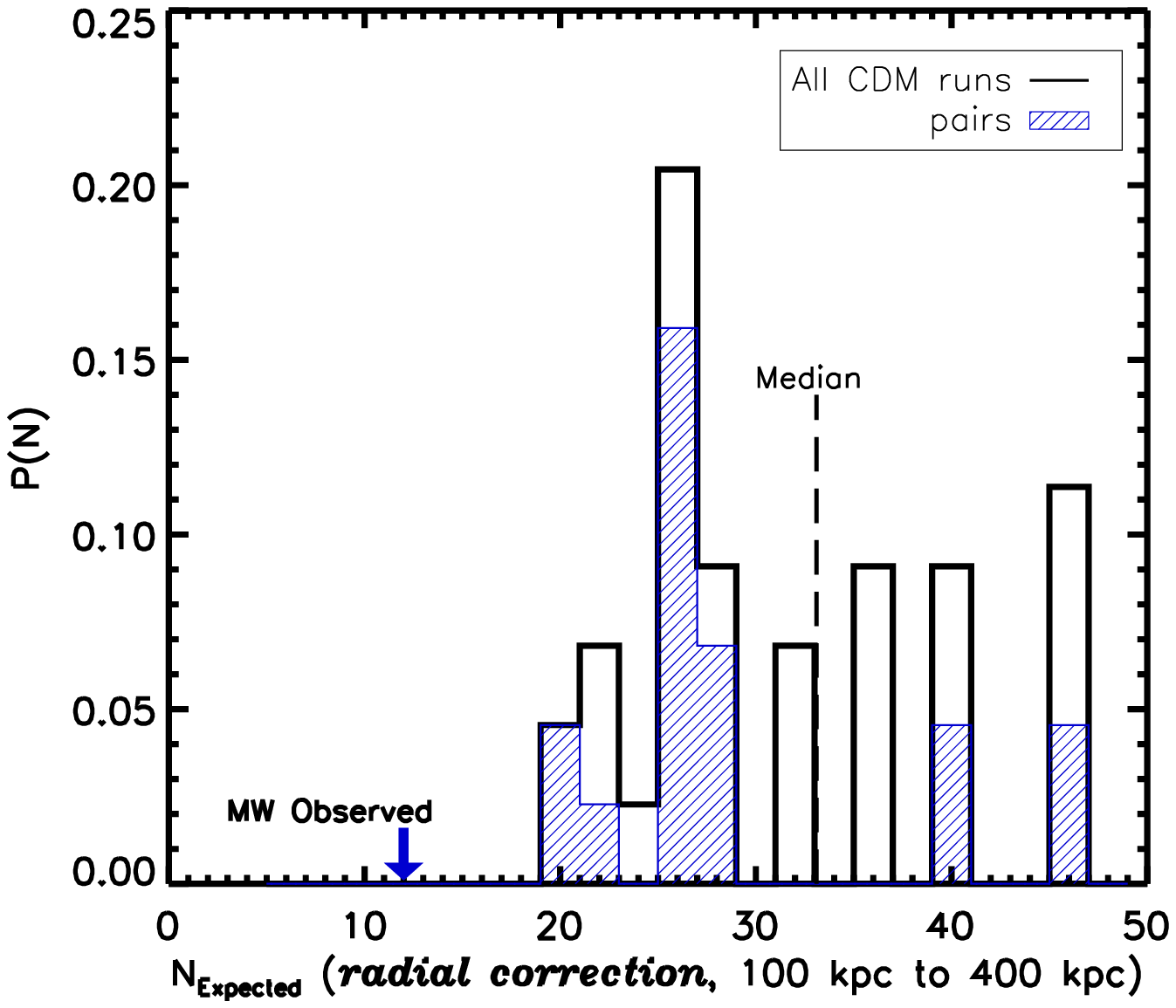}
 \caption{{\em Left:}
   The total number of Milky Way satellites expected over the whole sky
   given the observation that there are 6 bright satellites in the SDSS footprint 
 ($14555\,\deg^2$).  The calculation uses subhalos in our \lcdm\ simulations to correct for areal incompleteness,
 allowing for anisotropy.  The assumption is that only the SDSS area is known with
 confidence to be compete to systems with $L \sim 10^5 L_\odot$.       {\em Right:}  The total number
   of satellite galaxies expected with 400 kpc of the MW, estimated from the number of MW satellites known within
   100 kpc and then assuming that ratio of satellites within 400 kpc to those with 100 kpc matches the
    the expectation for \lcdm\ subhalos (see text).  The shaded 
   regions represent the contribution due to the
   paired ``Local Group-like" simulations.  The blue arrows indicate the
 number of known MW satellites brighter than $10^5 L_\odot$.
 \label{fig:correction}
}
\end{figure*}
%%%%%%%%%%%%%%%%%%%%%%%%%%%%%%%%%%%%%%%%%%%%%%%%%%%%%%%%%%%%%% 

Of the bright MW satellites, there are 6 that lie inside of the SDSS footprint.  We can be confident that our census is complete within
this region.  
Given this, we can provide a rough estimate for a plausible number of bright dwarfs that could exist outside this footprint  using our simulations.    Specifically, for each of our 44  systems,
we count subhalos in randomly-oriented angular cones the size of the DR8 footprint ($14555\,\deg^2$) with a depth of 400 kpc.  We 
determine a $\vpeak$ cut that yields 6 subhalos within this region.  We then determine the total number of subhalos over the whole sky
with $\vpeak$ higher than the cut.  This gives the total expected number of bright satellites within a spherical region of 400 kpc,
$N_{\rm expected}$, based only on the count within an SDSS area.

 For each host halo, we perform $10^4$ random realizations of mock pointings  to
 derive an $N_{\rm expected}$ distribution based on our full sample.  The results are shown in
Figure~\ref{fig:correction}.  The median prediction is 18 bright satellites over the whole sky.  This can be
compared to the 12 known bright MW satellites, so it is not implausible to think that $\sim 6$ undiscovered bright satellites of the Milky Way could be out there.
In fact, the distribution extends to at total $\sim 40$
satellites, which would correspond to a total of $\sim 28$ undiscovered satellites.

How many satellites would need to be discovered beyond 100 kpc if the MW
satellite distribution were to match the expectations of our \lcdm\ simulations?
The Milky Way has 7 bright satellites within 100 kpc, so for each of our
simulated halos, the value of $B_{400}$ can be used to compute the total number
of bright MW satellites expected based on the known number within 100 kpc:
\begin{equation}
N_{\rm expected} = N_{\rm MW}(< 100\,\rm kpc) \times B_{400} = 7 \, B_{400}\, .
\end{equation}
The right panel of Figure~\ref{fig:correction} shows that this predicts that
there would need to be a total of 20 to 40 bright MW satellites, substantially
more than the 12 that have been discovered so far.  The implication is that
there would need to be between $\sim 8-28$ new satellites beyond 100 kpc in
order for the Milky Way to fall within the range of our simulations.
Interestingly, this number overlaps with the allowed range from the angular
coverage correction discussed above (and in the left panel of
Figure~\ref{fig:correction}).

To have escaped
previous detection, a bright satellite would likely need to have a large half-light radius
($\sim 1$ kpc) and low surface brightness. Canes Venatici I, the bright
satellite discovered in SDSS, does indeed have such characteristics: its central
surface brightness is $\mu=27.1\,{\rm mag/arcsec^2}$ and its projected half
light radius is 560 pc \citep{Zucker_06b, McConnachie_12}.

\section{Discussion and Conclusions}
The results of Section~\ref{sec:results} show that the satellites of Milky Way
and M31 have very different radial distributions. Within 100 kpc of their
respective hosts, the $L > 10^5\, \lsun$ satellite galaxy populations of the two
galaxies have similar radial distributions and luminosity functions, whereas
between 100 and 400 kpc, M31 has approximately four times as many satellites as
the Milky Way. While \lcdm\ simulations predict substantial scatter in the
radial distribution of subhalos, the MW's satellite population is more
concentrated than \textit{all} of the simulations we have investigated.  This
result holds for reasonable associations between subhalos and galaxies, though
extreme subsets of subhalos, chosen on formation time, can alter this conclusion
(see below).  M31, on the other hand, looks fairly typical when compared to the
simulations.

It is possible that the Milky Way is simply an outlier in terms of the spatial
distribution of its satellites - though it would need to be extreme enough to be
rarer than one out of the 44 systems we have simulated.  A more intriguing
possibility is that the census of Milky Way satellites is incomplete not just
for the faintest dwarfs, but also for brighter systems ($L \sim
10^{5}\,\lsun$). The discovery of Canes Venatici I by \citet{Zucker_06b} in SDSS
data indicates that this is indeed a distinct possibility. Using the
distribution of subhalos from our simulations, we might expect 8-20 additional
satellites if the radial distribution of the MW satellites is similar to those
of subhalos in our simulations.  We showed that the angular anisotropy of
subhalos in our simulations allow for the possibility of 8-20 additional
satellites in the area of the sky not yet surveyed for new dwarfs by the SDSS.

If the  shape of the radial distributions of subhalos is a template for the radial
  distribution of satellite galaxies about their host, then the Milky Way almost certainly has
  new   bright satellites yet to be discovered.  This would be consistent with the expectations
  of hydrodynamical simulations of \citet{Bovill_11} who have argued that there should be more 
  bright satellites around the Milky Way based on their models. While the simulations we have
used for our comparison are dissipationless, including baryonic physics is
likely to exacerbate the discrepancy between the predicted and observed
distribution of satellites in the Milky Way, as the MW disk can deplete
substructure that pass nearby \citep{donghia_10}.

One way to increase the predicted concentration of the MW satellite population,
and therefore to bring the MW into better agreement with dissipationless \lcdm\
simulations, is to associate the Milky Way's bright satellites with its earliest
forming subhalos having $\vpeak > 10\,\kms$. Such an association also predicts
that the Milky Way's bright dwarf spheroidals (1) formed in halos with $\vpeak
\lesssim 30\,\kms$ and (2) reside in halos that currently have $\vmax \lesssim
20 \,\kms$. This scenario would be in agreement with the values derived by
\citet{bk_11} based on kinematics of the MW's bright satellites. In this case,
there would be 20-40 dark matter satellites of the Milky Way that are more
massive but less luminous than its bright dwarf spheroidal galaxies. This
scenario is highly unnatural in current models of galaxy formation.

Although we do not see a significant difference in the subhalo radial
distributions between paired and non-paired host halos, we can not rule out the
possibility that the evolutions of the Milky Way and M31 are coupled in some
way. The asymmetrical 3D distribution of M31's satellites \citep{Conn_13} may
hint at such a connection.  We are currently using the ELVIS suite to evaluate
the frequency of such ``great plane'' configurations in \lcdm\ simulations.

\section*{Acknowledgements}

Support for this work was provided  by NASA through a grant for program AR-12836
from the Space Telescope Science Institute (STScI), which is operated by the
Association of Universities for Research in Astronomy (AURA), Inc., under NASA
contract NAS5-26555. MB-K acknowledges support from the Southern California
Center for Galaxy Evolution and the UC High Performance Astro-Computing Center,
multi-campus research programs funded by the University of California Office of
Research.   Initial calculations leading to this work was partially supported by HST-GO-12273.03-A.
BY was supported by NSF grant AST-1009973.  SG-K was supported by NSF grant AST-1009999.

\newpage
\section*{Appendix}
%%%%%%%%%%%%%%%%%%%%%%%%%%%%%%%%%%%%%%%%%%%%%%%%%%%%%%%%%%%%%%
\begin{figure}
\hspace{-16mm}			
 \includegraphics[width =  0.65\textwidth]{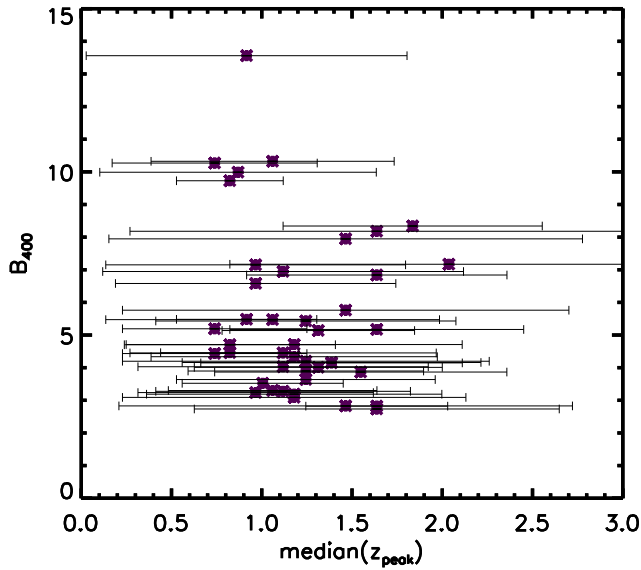} 
 \caption{Subhalo concentration distributions ($B_{400}$) plotted as a function of
   $z_{\rm{peak}}$, the median redshift at which $\vpeak$ is attained
   for each subhalo system.   This is analogous to comparing $B_{400}$ to the
   median infall time of each subhalo set.  The error bars show the $1\sigma$ 
   spread around the median value of $z_{\rm peak}$ for each subhalo set.  There is no evident correlation, 
   and we are also unable to find a correlation with any property of the main halo 
   (e.g., concentration, mass, or formation time.)}
\label{fig:apeak}
\end{figure}
%%%%%%%%%%%%%%%%%%%%%%%%%%%%%%%%%%%%%%%%%%%%%%%%%%%%%%%%%%%%%% 

It is reasonable to expect the concentration factor, $B_{400}$, to be correlated
to with some property of host halo. For example, the concentrations of dark
matter halos are known to be correlated with formation redshifts \citep{nfw1997,
  Bullock_01}.  We have searched for relationships between $B_{400}$ and several
properties of the host halo, including virial mass, virial radius, $\rmax$,
$\vmax$, and $M(<\rmax)$, but were unable to find any
correlations. Figure~\ref{fig:apeak} illustrates one such null result: $B_{400}$
is plotted as a function of the median (asterisks) and 68\% spread (error bars)
in $a_{\rm peak}$, the expansion factor at which a subhalo reaches $\vpeak$, for
each ELVIS halo. While the spread in $a_{\rm peak}$ is large for each system,
there is no correlation with $B_{400}$.

%%%%%%%%%%%%%%%%%%%%%%%%%%%%%%%%%%%%%%%%%%%%%%%%%%%%%%%%%%%%%% 
\begin{figure}
\hspace{-16mm}
 \includegraphics[width =  0.65\textwidth]{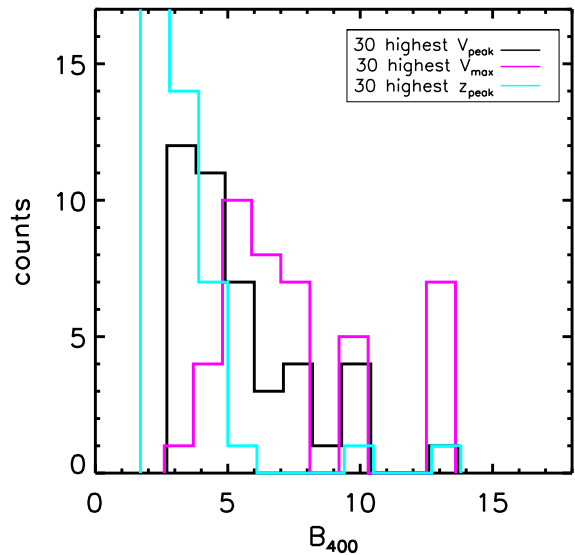} 
 \caption{A histogram of $B_{400}$ values for all 44 host halos computed by
   selecting the subhalos with the highest $\vpeak$ (black), $\vmax$ (magenta)
   and $z_{\rm peak}$ (cyan).  The range and distribution of $B_{400}$ depend
   somewhat on how subhalos are selected.  Our selection of subhalos with the 30
   highest $\vpeak$ predicts somewhat more concentrated distributions of
   satellites than does the selection based on $\vmax$. The most concentrated
   distributions are found when selecting based on the earliest forming (highest
   $z_{\rm peak}$) subhalos.}
\label{fig:b400_comp}
\end{figure}
%%%%%%%%%%%%%%%%%%%%%%%%%%%%%%%%%%%%%%%%%%%%%%%%%%%%%%%%%%%%%%

\begin{figure*}
 \includegraphics[width = 0.55\textwidth]{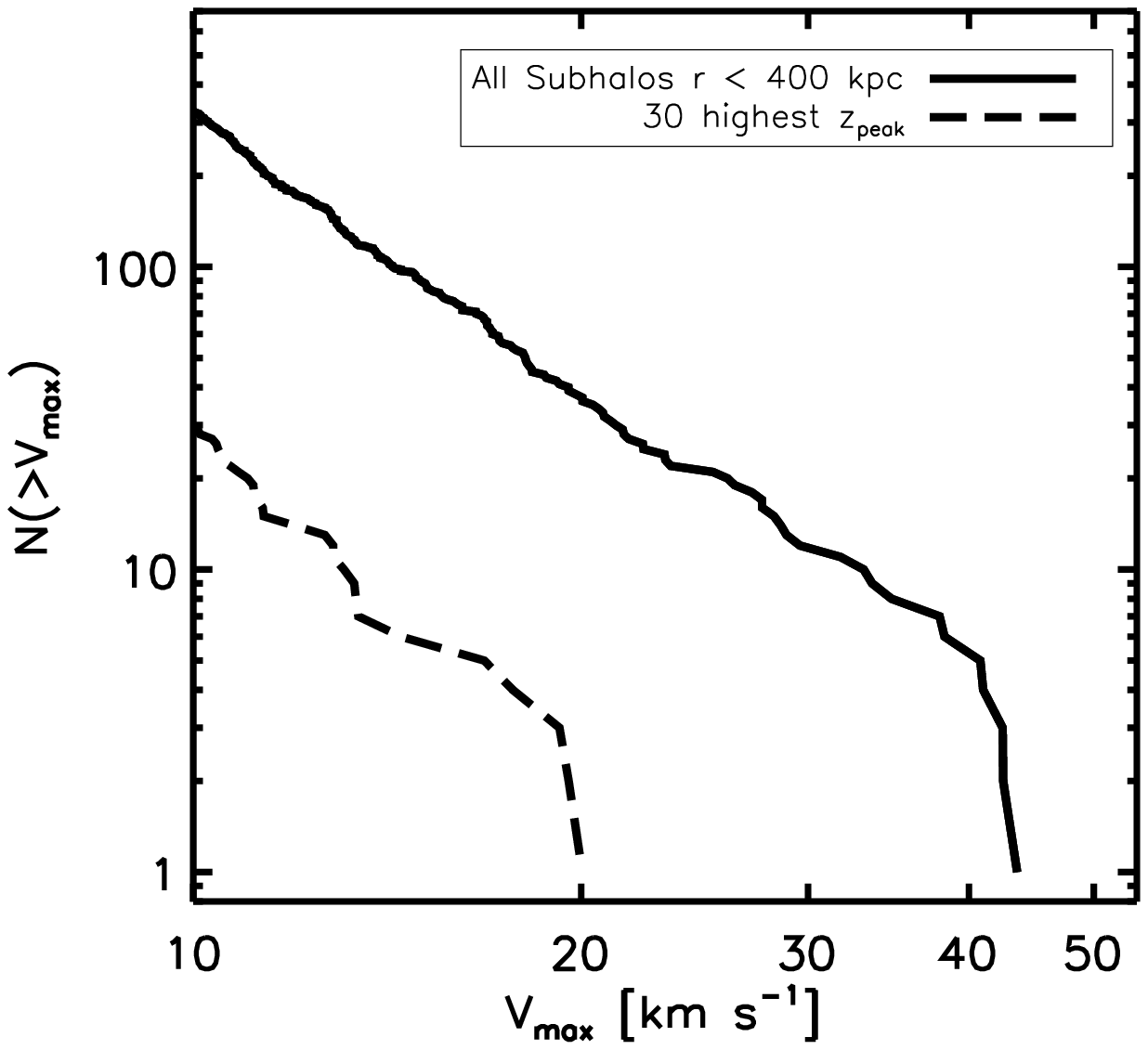}
 \hspace{-24mm}
 \includegraphics[width = 0.55\textwidth]{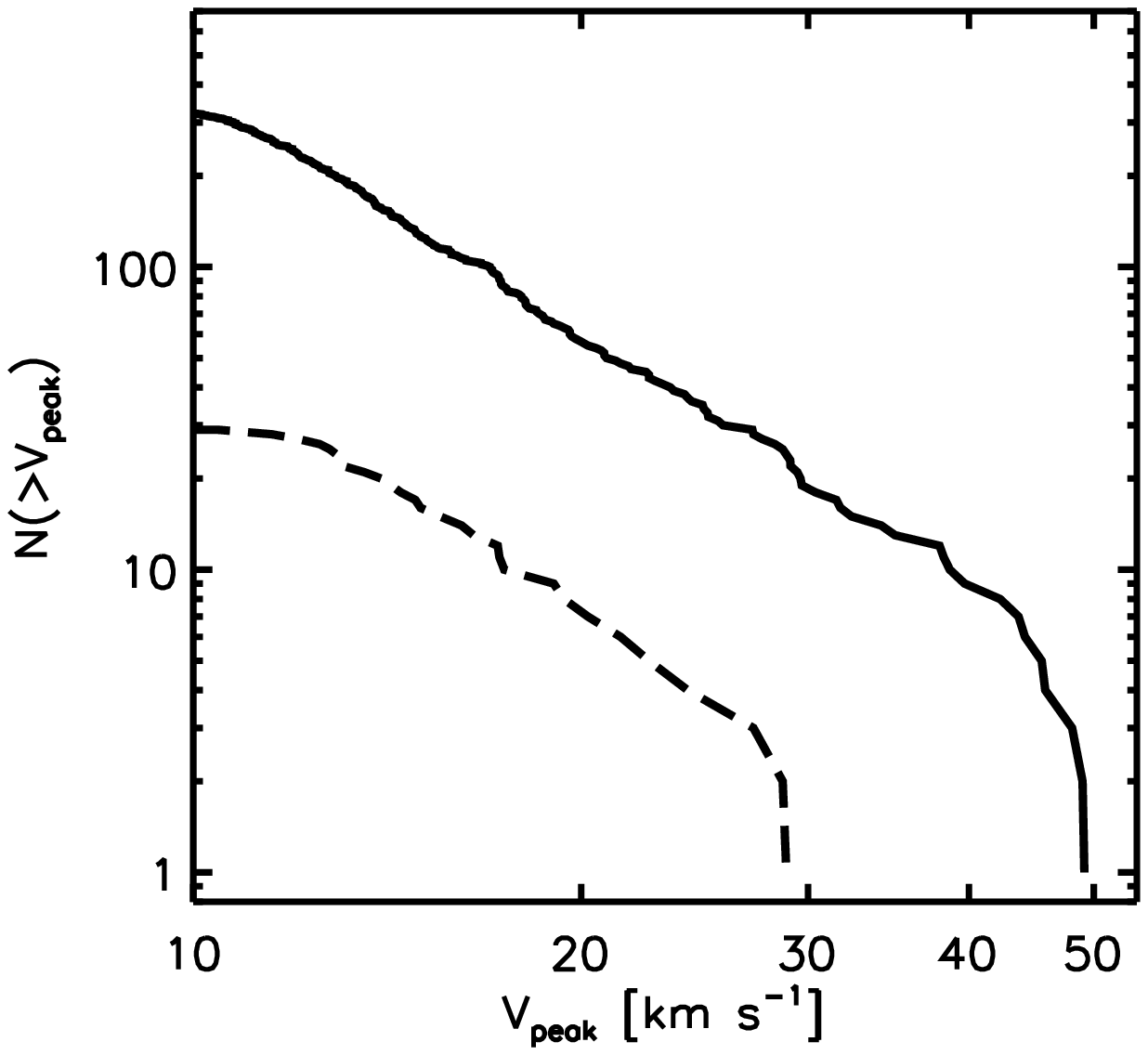}
 \caption{Cumulative counts of subhalos as a function of $\vmax$ (left) and
   $\vpeak$ (right) in a representative simulated halo. The solid lines
   correspond to all subhalos within 400 kpc of host halo, while the dashed
   lines represent a subset of those halos with the highest values of $z_{\rm
     peak}$, the redshift at which the subhalo reached the $\vpeak$. The
   $\sim10$ most massive subhalos have $\vpeak \gtrsim 40\,\kms$ or $\vmax
   \gtrsim 30 \,\kms$, whereas the $\sim$10 largest early forming subhalos have
   $\vpeak \gtrsim 18\,\kms$ or $\vmax \gtrsim 12 \,\kms$. 
   \label{fig:vfunc_comp}}
\end{figure*}

Figure~\ref{fig:b400_comp} explores a few different ideas for associating
subhalos with bright satellites and the effect of that selection on their radial
distributions.  Specifically, for each of our 44 hosts we plot the implied
$B_{400}$ (puffiness) parameter (Equation 2) derived for three different choices
of subhalo selection: the thirty highest $\vpeak$ (black, fiducial); the 30
highest $\vmax$ (magenta); and the 30 highest $z_{\rm peak}$ (cyan; defined as
the redshift where $\vpeak$ was reached).  We see that by selecting on $z_{\rm
  peak}$ we produce distributions that are not as puffy as our fiducial choice,
and marginally consistent with the $B_{400}$ of the MW, which is 1.7.  However,
subhalos chosen in this manner are an odd subset in terms of their mass.
Figure~\ref{fig:vfunc_comp} shows the $\vpeak$ and $\vmax$ mass functions of
one example system selected in this manner (dashed) compared to all of the
subhalos associated with this host (solid).  It would demand a large number of
$\vpeak > 30 \kms$ and $\vmax > 20 \kms$ subhalos be dark.  Such a model would
be very difficult to understand from a galaxy formation perspective, as there is
no obvious mechanism for allowing galaxies to form in halos of $30\,\kms$ while
suppressing galaxy formation in halos of $30-50\,\kms$.  Moreover, the choice
amounts to selecting subhalos that reached their peak $\vmax$ value at redshifts
between 2 and 4, well later than the epoch of reionization, which might offer
more obvious channels for galaxy suppression mechanisms.

%%%%%%%%%%%%%%%%%%%%%%%%%%%%%%%%%%%%%%%%%%%%%%%%%%%%%%%%%%%%%%
%%%%%%%%%%%%%%%%%%%%%%%%%%%%%%%%%%%%%%%%%%%%%%%%%%%%%%%%%%%%%% 

\label{lastpage}
\end{document}